\documentclass[twocolumn,showpacs,preprintnumbers,amsmath,amssymb]{revtex4}
\usepackage{graphicx}
\usepackage{dcolumn}
\usepackage{bm}
\usepackage{enumerate}

\begin{document}

\title{Constructing a counterexample to the black hole complementarity}

\author{Dong-han Yeom}
\email{innocent@muon.kaist.ac.kr}
\author{Heeseung Zoe}
\email{hszoe@muon.kaist.ac.kr}
\affiliation{Department of Physics, KAIST, Daejeon, 305-701, South Korea}
\date{\today}

\begin{abstract}
We propose a regular black hole whose inside generates a de Sitter space
and then is finally frustrated into a singularity.
It is a modified model which was suggested originally by Frolov, Markov, and Mukhanov.
In our model, we could adjust a regular black hole so that its period before going into the extreme state is much longer than the information retention time.
During this period an observer could exist who observes the information of the Hawking radiation,
falls freely into the regular center of the black hole,
and finally meets the free-falling information again.
The existence of such an observer implies that the complementary view may not be consistent with a regular black hole,
and therefore, is not appropriate as a generic principle of black hole physics.
\end{abstract}
\maketitle

\section{\label{sec:intro}Introduction}

The black hole information paradox is one of the most important and difficult problems in modern physics \cite{inforpara}.
Resolutions have been proposed by various authors with different motivations (e.g., \cite{Susskind:2002ri}\cite{Hawking:2005kf}\cite{Ashtekar:2005cj}).
Remarkably, the black hole complementarity principle \cite{complementarity}\cite{nonlocal}\cite{inforretention} cooperates well with
string theory based on holography (e.g., \cite{Susskind:1994vu}\cite{Callan:1996dv}\cite{Maldacena:1997re}).

According to the complementary view of a black hole, information about matter which falls into the black hole is actually copied near the event horizon.
The asymptotic observer observes that the information resides near the event horizon, and the information will reemit in the form of Hawking radiation.
On the other hand, the free-falling observer who goes beyond the horizon, can always observe the original information,
and the information is not affected by Hawking radiation.
In fact, neither of these two kinds of different observations are permitted by the no cloning theorem,
but if no observer has access to both pieces of information, in other words,
if the asymptotic and the free-falling observer cannot communicate forever, then there will essentially be no problem.

However, a problem remains in terms of how to make a connection between the two copied pieces of information of the complementary observers.
To archive the complementary view, we may require that ``nonlocality", as a fundamental ingredient of quantum gravity, should be realized \cite{nonlocal},
or there may be a nonunitary collapse near the singularity \cite{hmproposal}.
There are some proposals which are closely related to the black hole complementarity, but no commonly accepted conclusion on the issue seems to exist.

If the complementarity is true, then we at least have a nice picture of the information flow for an asymptotic observer.
But, what if an outer observer sees the Hawking radiation and free-falls into the black hole (we will call this \textit{a duplication experiment})?
Then the observer may verify whether or not the information duplication actually happened, and the complementarity principle may be falsified.
The inventors of the complementarity argue that we can circumvent this problem as follows: since the outer observer must wait by the information retention time \cite{Page},
and since this time scale is quite long, it will be almost impossible for the outer observer to meet the free-falling information before touching the singularity \cite{inforretention}.

In this paper, we reconsider the complementarity in the context of a regular black hole.
In Sec. \ref{sec:fmm}, the Frolov, Markov, and Mukhanov's model of a regular black hole is reviewed.
Physical initial conditions are also introduced for the duplication experiment.
In Sec. \ref{sec:cauin}, the causal structure of our model is explained,
and the information which flows in and out around the horizon is discussed.
In Sec. \ref{sec:inner}, the penetrability of the inner horizon and the safety of the inside structure are discussed.
In Sec. \ref{sec:duplication}, the duplication experiment is shown to be realized,
and in Sec. \ref{sec:dis}, its implications to the black hole complementarity and holography are discussed.

\section{\label{sec:fmm}Frolov, Markov, and \protect\\ Mukhanov's model}

There are many well-known models of regular black holes \cite{regular}\cite{Hayward:2005gi}.
We will use the model of Frolov, Markov, and Mukhanov \cite{Frolov:1988vj}.
If there is a local false vacuum and we push some matter to it, then there will be a black hole without singularity
since the inside of the black hole becomes a de Sitter space.
However, in order to paste two different vacua, we may need a transition layer
which would be approximated by a thin massive shell \cite{Israel:1966rt}.
Originally, this model was used to replace the space around the singularity with a regular de Sitter space,
using a principle known as the limiting curvature hypothesis \cite{Frolov:1988vj}\cite{lch}.
However, we do not appeal to this hypothesis and use only its metric.
So the local false vacuum needs not be as small as the Planck size.

The metric and the energy-momentum tensor of the massive shell are as follows:
\begin{eqnarray} \label{metric}
ds^{2} = -\left(1-\frac{2m(r,l)}{r}\right) dt^{2} \nonumber \\
+\left(1-\frac{2m(r,l)}{r}\right)^{-1}dr^{2}+r^{2}d\Omega^{2}.
\end{eqnarray}
The mass function $m(r,l)$ becomes $m\theta(r-r_{0})+(r^{3}/2l^{2})\theta(r_{0}-r)$,
where $l=(\Lambda/3)^{-1/2}$ is the Hubble scale parameter and $r_{0}=(12/\alpha)^{1/6}(2m/l)^{1/3}l$ is the radius of the false vacuum boundary
(we can choose the value of $\alpha$ as a free parameter).
Then, one can easily check that (if we choose $\alpha = 12$) the metric gives the outer horizon ($r_{+}=2m$)
and the inner horizon ($r_{-}=l$), and usually $r_{-}<r_{0}<r_{+}$ holds as long as $l\ll m$.
If $r<r_{0}$, the metric is exactly the same as a de Sitter space, and, otherwise, it is exactly the same as a Schwarzschild black hole.
We can calculate a proper mass shell condition \cite{Frolov:1988vj}:
\begin{equation} \label{energy-momentum tensor}
S^{\mu}_{\nu} = \textrm{diag}\left(\frac{\lambda}{4\pi},0,\frac{\kappa+\lambda}{8\pi},\frac{\kappa+\lambda}{8\pi}\right),
\end{equation}
where
\begin{eqnarray}
 \kappa  & = & \frac{r_{0}}{l^{2}}\left[\left(\frac{r_{0}}{l}\right)^{2}-1\right]^{-1/2}+ \frac{m}{r_{0}^{2}}\left[\frac{2m}{r_{0}}-1\right]^{-1/2}, \\
 \lambda & = & \frac{1}{r_{0}}\left[\left(\frac{r_{0}}{l}\right)^2-1\right]^{1/2}- \frac{1}{r_{0}}\left[\frac{2m}{r_{0}}-1\right]^{1/2}.
\end{eqnarray}
This massive shell can be constructed by ordinary scalar (matter) fields \cite{wall}\cite{Blau:1986cw}.

One problem of the initial condition is the origin of the local false vacuum.
One may guess that, since the false vacuum inflates and it will deflate to the past direction, we may start the initial singularity \cite{Farhi:1986ty};
however, this happens only for unbuildable states \cite{Freivogel:2005qh},
and since we do not consider the exponentially expanding vacuum (our vacuum will collapse to the singularity),
we can think that we start from the buildable vacuum prepared in unitary processes.
To prepare our local false vacuum, we may need to assume that the background is a kind of de Sitter space for the energy conservation problem.
Although we assume this, as long as the false vacuum is almost a true vacuum,
our metric form will not be so different around the black hole radius (see Appendix B).

Let us assume that the change of mass $m$ or parameter $l$ is sufficiently slow, and then we can use the metric form as Vaidya \cite{Frolov:1988vj}.
One may notice that, if there is an initial local vacuum and at an ideal time, if we push some critical mass($m_{*}=l/2$) to the vacuum,
there will be a black hole with one horizon at $r_{+}=r_{-}=r_{0}=l$ since the outer and the inner horizons are the same.
The fact that we can assume the metric structure before this time is supported by the stability of G-lumps \cite{Dymnikova:2007gx}.
The geometry of this black hole is described as a junction between a Friedmann space and a de Sitter space.
By adjusting $\alpha$, we initiate the regular black hole with no mass shell \cite{Frolov:1988vj}.
As the mass of the black hole grows, two horizons will be separated, and $r_{0}$ will be located between two regions.

After the mass supply ends, the Hawking radiation becomes important.
We can easily calculate the Hawking temperature of some regular black holes (See Appendix A.)
There are two potential problems:
$m(r,l)$ is not a $C^{\infty}$ function and
the value $\kappa$ becomes large as the black hole approaches becoming an extreme black hole.
However, the thin shell approximation would not be valid
if the transition layer was comparable to the length difference between two horizons, as they approach the extreme limit.
So, the approximation of the transition layer should be modified so that the metric and field contents are regularized around the extreme limit
(a possible modified version of a regular black hole is described in \cite{Hayward:2005gi}).
Then, we know that our calculation of Hawking temperature is still valid
and conclude that it would become $0$ as the black hole approaches the extreme limit.
Since the evaporation process is sufficiently slow, because of the stability of the mass shell \cite{Balbinot:1990} for small perturbations,
the vacuum will not collapse to singularity, and we can assume the metric structure (i.e., Vaidya type structure) until nearly the extreme limit.
In any case, there is no problem to penetrate the shell along the radial direction, and its energy density will become zero around the extreme limit.

Finally, the field of the false vacuum rolls down to another false vacuum.
In this stage, we assume that the perturbation is large enough to form a singularity by the collapse of the mass shell,
so the mass shell must collapse and form a Schwarzschild black hole.
However, the internal matter will not form a singularity until the mass shell collapses because the inside is still a false vacuum
(for detailed analysis, see \cite{Balbinot:1990} and Appendix B).

\begin{figure}
\begin{center}
\includegraphics[scale=0.85]{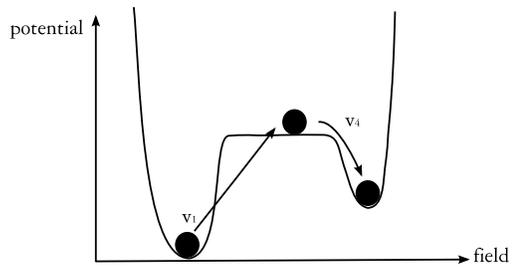}
\caption{\label{fig:vacuum}The scalar field of a local false vacuum changes. At time $v_{1}$, it is generated.
At time $v_{4}$, it starts to decay.}
\end{center}
\end{figure}

\section{\label{sec:cauin}Causal structure and information flow}

\begin{figure}
\begin{center}
\includegraphics[scale=0.5]{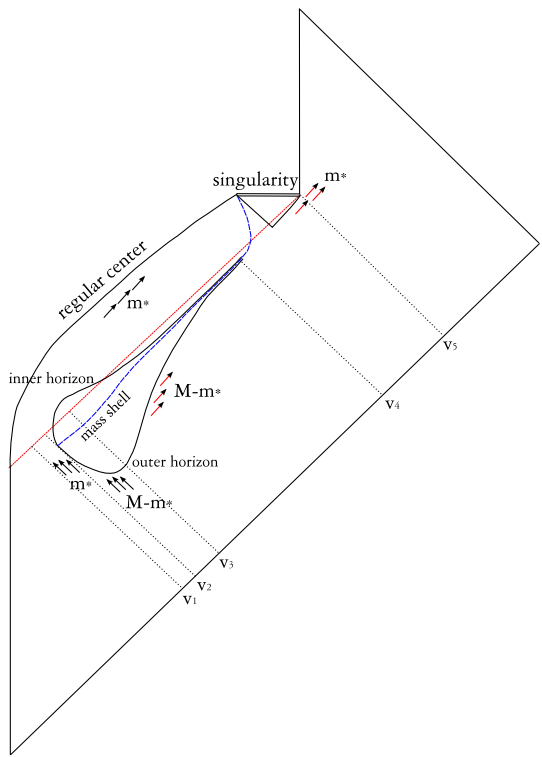}
\caption{\label{fig:penrose}The Penrose diagram. This shows the inner horizon and the outer horizon. The dashed curve is the mass shell.
The left curve means the regular center, and we regard it as a timelike curve.
The arrows are the in-falling matter and out-going matter (i.e., the Hawking radiation).
Eventually, one can see the event horizon as a thin null line.}
\end{center}
\end{figure}

\begin{figure}
\begin{center}
\includegraphics[scale=0.75]{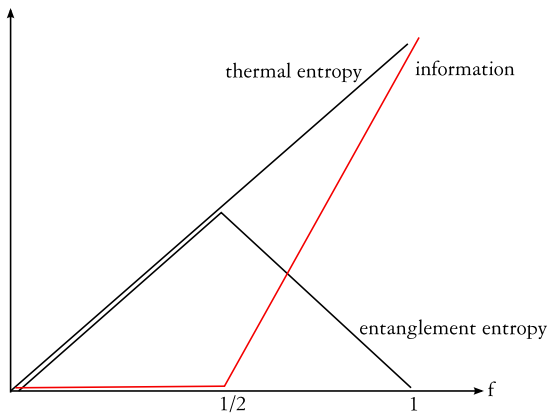}
\caption{\label{fig:information}Flow of information \cite{inforretention}\cite{Page}.}
\end{center}
\end{figure}

We can draw the whole causal diagram of our model (Fig. \ref{fig:penrose}).
The advanced time $v$ can be used as a time parameter in the Vaidya metric.
Until $v_{1}$, the space-time is flat. Around $v_{1}$, a false vacuum is generated (Fig. \ref{fig:vacuum}).
Between $v_{1}$ and $v_{2}$, the critical mass falls into the black hole, and there is no horizon.
After $v_2$, horizons are generated, and a mass shell is also generated between two horizons.
As mass falls, the outer horizon grows in a spacelike direction and the inner horizon in a timelike direction.
After $v_{3}$, the mass flow ends, and the Hawking radiation becomes important;
the outer horizon moves in a timelike direction, and the inner horizon in a spacelike direction
(on the dynamics of local horizons, see \cite{Ashtekar:2004cn}).
After $v_{4}$, or after two horizons approach, the scalar field decays, and the mass shell collapses to form a singularity.
The regular center and the mass shell will approach space-like singularity, and the regular center must be connected smoothly in a timelike direction.
This will form the left boundary of the Penrose diagram.
After $v_{4}$, since there is no significant effect on the outside observer, the geometry is exactly like a Schwarzschild black hole.
Because parameter $l$ increased, although the mass shell collapses, the outer and the inner horizons are outside of the shell.
Thus, one may guess that two horizons disappear after $v_{4}$ and the mass shell will form another apparent horizon as it collapses \cite{Ashtekar:2004cn}.
After $v_{5}$, the evaporation ends, and the final spacetime becomes flat again.
Finally, we can draw the event horizon of this black hole (each step of the causal diagram is consistent with \cite{Frolov:1988vj},
and one can compare and find some differences in \cite{Hayward:2005gi}).

One may guess that the left boundary must be a straight line from bottom to top, but it is still enough to understand the essential behavior; furthermore, we can simply modify it to a straight line.

We assume that the time evolution of the black hole is unitary.
Then, it is known that, its entanglement entropy starts from $0$ at $v_{3}$ and reaches the maximum as the thermal entropy
(or, equivalently, its area) of the black hole becomes half of its original value \cite{Page}.
If $l\ll M$, where $M$ is the maximum mass of the black hole, then the half point will be located between $v_{3}$ and $v_{4}$
and the entanglement entropy will approach $0$ as the black hole evaporates since we assume the unitarity.
However, we know that the thermal entropy will increase as the area of the black hole decreases.

Therefore, if we choose $f$, the fraction of the total degrees of freedom contained outside as an $x$-axis parameter,
then the entanglement entropy and the thermal entropy behave as in Fig. \ref{fig:information}, and we know the information from the definition \cite{inforretention}\cite{Page}:
\begin{equation} \label{information}
I = S_{thermal} - S_{entanglement}.
\end{equation}
So one finds the information retention time.
The information retention time will be between $v_{3}$ and $v_{4}$.
After the information retention time, the entanglement entropy decreases monotonically
and the thermal entropy increases monotonically.
Therefore, \textit{if the escaped mass after the information retention time is not negligible, it must contain some information.}
From this point forward, we can regard the flow of mass as equivalent to the flow of information.

Now we estimate the flow of mass (Fig. \ref{fig:penrose}).
Between $v_{1}$ and $v_{2}$, we push critical mass $m_{*}\approx l/2$.
As long as $l$ is not too small, the critical mass is not negligible.
After $v_{2}$, we push $M-m_{*}$ until $v_{3}$.
From $v_{3}$ to $v_{4}$, since the black hole becomes extreme, $M-m_{*}$ escapes by the Hawking radiation.
After $v_{4}$, the mass must remain $l/2$ inside of the black hole.
Finally, after $v_{5}$, $l/2$ will escape.

\section{\label{sec:inner}On instability of inner horizon}

One possible problem is the instability of the inner horizon.
This problem was suggested in the context of the cosmic censorship of charged black holes \cite{Simpson:1973ua}.
This instability was identified with the effect known as mass inflation \cite{Poisson:1990eh},
which induces singularity along the inner horizon since the curvature becomes infinite
(this singularity was regarded as a boundary condition or important boundary by some authors \cite{hmproposal}\cite{Ge:2005bn}\cite{Thorlacius:2006tf}).
However, the inner horizon singularity is weak enough and does not imply the end of space-time, like the Schwarzschild singularity \cite{Ori:1991},
so although the penetration is difficult, the inside structure may be safe.

We have to consider two problems: one is \textit{whether the inner horizon can be penetrable or not}
and the other is \textit{whether the inside of the inner horizon can be safe or not}.
Recently, the authors performed numerical calculations for dynamical charged black holes, and we will report on some of the results \cite{HHSY}.

For the first problem, we notice that the mass function around the inner horizon $m(u,v) \sim \exp{\kappa_{i}(u+v)}$ (of course, this behavior will be common for the inner horizon of a regular black hole \cite{Poisson:1997my}) becomes infinite only for
$u\rightarrow\infty$ or $v\rightarrow\infty$ limit,
where $u$ and $v$ are coordinate variables of the double null coordinate, and $\kappa_{i}$ is the surface gravity of the inner horizon.
If we turn on the Hawking radiation, all locations of the Penrose diagram are accessible in finite $u$ and $v$; thus, the mass function is finite everywhere.
Therefore, there is no curvature singularity in the classical sense, and a field or matter will be penetrable \cite{HHSY}.

The second problem is whether or not the inner horizon collapses and forms a strong singularity due to some perturbations.
However, we know that the inner horizon is regular and penetrable from the previous remarks.
If it is penetrable, as long as the perturbation is small enough, it will not destroy the inside structure.
Thus, as long as we push matter or signals slowly, we can trust the metric structure everywhere (at least, qualitatively).
(This conclusion is also supported by some stability arguments \cite{Dymnikova:2007gx}\cite{Balbinot:1990}.)

One potential problem is that the mass function or curvature function becomes large (possibly greater than the Planck scale) for charged black holes.
This problem may occur in regular black holes, but the situation will be better than charged black holes since there is no strong singularity inside of regular black holes.
Moreover, we noticed that, as we choose a large number of massless fields, we can tune the Planck cutoff scale to be larger and larger \cite{HHSY};
in this limit, we can trust the entire region with a semiclassical description except the classical singularity.

In conclusion, we found physically possible conditions in which we can trust the semiclassical description of our model.
Note that, since the singularity only happens at the final stage of the black hole, the Horowitz-Maldacena's proposal \cite{hmproposal} cannot work before $v_{4}$.

\section{\label{sec:duplication}The duplication experiment}

\begin{figure}
\begin{center}
\includegraphics[scale=0.5]{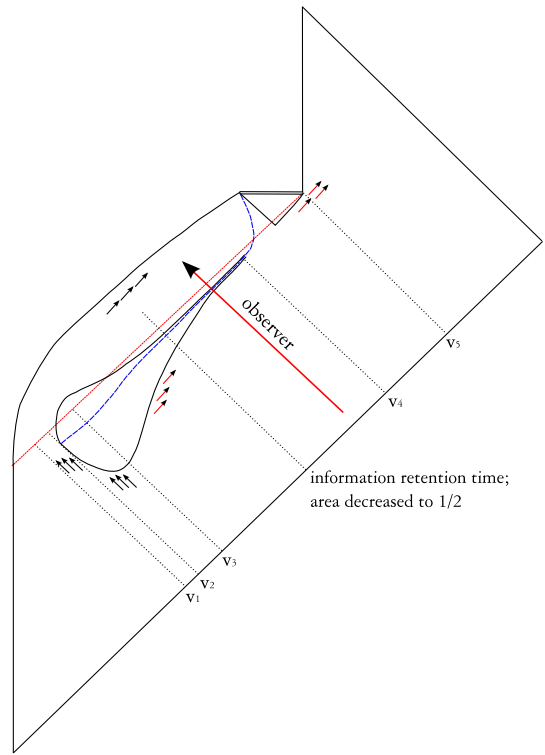}
\caption{\label{fig:duplication}The duplication experiment.}
\end{center}
\end{figure}

Now, we are ready to perform the duplication experiment (Fig. \ref{fig:duplication}).

We use the free-falling observer along the null direction between the information retention time and $v_{4}$.
The observer can observe enough information from the Hawking radiation,
and the penetration seems to be possible, as we discussed previously.
Thus, the observer can compare the Hawking radiation with almost all of the free-falling information.
The existence of an observer performing this duplication experiment invalidates the no cloning theorem;
therefore, there is an observer who observes the violation of a natural law.
This implies the violation of the black hole complementarity.

One may suspect that, around the extreme limit, the information can escape from the shell, since the outer horizon crosses the shell. However, this is not a real problem in the duplication experiment for the following two reasons. First, information should escape before the outer horizon closely approaches the mass shell; thus, the duplication experiment will be possible even if all of the shell is inside of the outer horizon. Second, the mass shell has small mass compared to the initial mass $M$, and so it cannot contain enough information; thus, it will not be helpful to rescue the complementarity principle.

In \cite{Ge:2005bn}, the same gedanken experiment is proposed in a charged black hole,
but they did not fully considered dynamical cases; thus, the conclusion was unclear.
However, now we have a concrete model.
Of course, one can argue that maybe there are some incorrect assumptions in our model.
One point of concern is whether it may be possible to destroy the $m_{*}$ so $m_{*}$ experiences some Planck scale region before $v_{4}$.
Furthermore, one may suspect that the inner horizon never became penetrable.
However, according to our analysis \cite{HHSY}, the inner horizon can be regular and penetrable, as well as have low curvature, when we assume a large number of massless degrees of freedom.
If the assumption is not fundamentally impossible, our model can be meaningful to test the complementarity principle.

\section{\label{sec:dis}Discussion}

Although some fine-tuning is needed, it seems to be possible to construct the regular black hole of our model.
The existence of an observer performing the duplication experiment means that the black hole complementarity is not consistent to some extent.
It is now a proper step to ask the validity of the complementarity, which is to be ``operationally meaningful.''
In fact, the complementary view is an inevitable choice to protect the holographic principle and the unitarity of the quantum mechanics; however, we argue that our model would work as a counterexample.

One may say that our model is delicately fine-tuned to invalidate the black hole complementarity,
and the duplication experiment is successful only in a certain type of regular black holes.
Then, it would be fair to say that the complementarity is true \textit{effectively}
in facing various gravitational systems like black holes in general relativity.

This insight leads us to make a cautious remark on using the black hole complementarity on the inflationary measure problem \cite{holomeasure}.
The difficulty of finding a proper measure in a multiverse where eternal inflations take place has been discussed in many instances.
It is known that, by assuming the complementarity, we may suggest a better measure without those difficulties.
However, if the suggested measure assumes the complementarity, and the complementarity principle cannot be true for all situations,
the measure will possibly not work.

There are various interpretations of the holographic behavior of black hole entropy.
While the holographic principle of string theory implies that
the real information constructing the black hole is encoded on the horizon,
another interpretation of the holographic principle implies that the outer horizon \textit{looks like} a holographic screen since we cannot access beyond the horizon in a practical sense.
For example, loop quantum gravity provides the entropy formula by using this interpretation on the holography \cite{lqg}.
They (e.g., loop quantum gravity area) give ``operationally practical'' ways to define the accessible degrees of freedom to an asymptotic observer.
In this context, the holography does not need to be protected by the complementarity
and could be consistent with a naive expectation of general relativity near the black hole horizon.
Thus, our model may be able to cooperate with this rather weak version of the holographic principle.
However, the perspective of a dynamical observer is not clear,
and, of course, the information paradox puzzle should be resolved in this case
(see discussions in \cite{Ashtekar:2005cj}\cite{Ashtekar:2005qt} and also \cite{Vachaspati:2006ki}\cite{Hayward:2005ny}).

Therefore, the authors think that this gedanken experiment reveals the limitation of the complementarity principle.
It seems that although the string theory and the holographic principle may be fundamentally true, they must be modified within a certain limit.

\acknowledgments{
The authors would like to thank Ewan Stewart, Sean Hayward, and Alex Nielsen for their discussion and encouragement.
The authors also thank Irina Dymnikova for useful comments on G-lumps, and Dong-il Hwang and Sungwook Hong for insightful conversation.
This work was supported by BK21 and the Korea Research Foundation Grant funded by the Korean government (MOEHRD; KRF-2005-210-C000006, KRF-2007-C00164).
The authors would also like to thank the Korea Science and Engineering Foundation(KOSEF) for a grant funded by the Korean government (No. R01-2005-000-10404-0).
}

\appendix

\section{\label{appa}Hawking Temperature of a Regular Black Hole}

Let us begin with the following metric form:
\begin{equation} \label{regularmetric}
ds^{2}=r^{2}d\Omega^{2}+\frac{dr^{2}}{F(r)}-F(r) dt^{2},
\end{equation}
where $F(r)=1-2M(r)/r$ and $M(r)$ is a regular function. Then $F(r_{\pm})=0$ holds.

At first, the Hawking temperature may be proportional to the surface gravity of the outer horizon ($\kappa_{o}$).
For the spherically symmetric case, the surface gravity on the trapping horizon \cite{Ashtekar:2004cn}
is calculated by these authors (e.g.,\cite{Nielsen:2007ac}). The result is as follows:
\begin{equation} \label{surfacegravity}
\kappa_{o} =\frac{1}{4M(r_{+})}\left( 1-2M'(r_{+}) \right).
\end{equation}

Now we have to check whether or not the surface gravity is proportional to the Hawking temperature.
We prove this in two ways. (Note that the Hawking temperature must be related with local horizons \cite{Visser:2001kq}).

First, we use the Euclidean rotation method \cite{Zee:2003mt}.
If we use the Wick rotation on the metric (\ref{regularmetric}), since the topology changed, we can choose a new metric form:
\begin{equation} \label{wickrotation}
ds^{2}=R^{2}d\alpha^{2}+dR^{2}+r^{2}d\Omega^{2},
\end{equation}
where
\begin{equation} \label{R2pi}
R(2 \pi) = F(r)^{1/2} \beta,
\end{equation}
\begin{equation} \label{R}
R = \int ^{r}_{r_{+}} F(r')^{-1/2} dr',
\end{equation}
and $\beta$ is the period of the Wick rotated time. We can identify this as the inverse of the Hawking temperature.

Now we use the Taylor expansion near $r_{+}$ for $F(r)$. Since $F(r)$ vanishes at $r_{+}$, the result is
\begin{equation} \label{Fexpansion}
F = \left(\frac{2M(r_{+})}{r_{+}^{2}}-\frac{2M'(r_{+})}{r_{+}}\right)(r-r_{+}) + O[(r-r_{+})^{2}].
\end{equation}
And after changing $r_{+}$ to $2M(r_{+})$, we obtain $2\kappa_{o} (r-r_{+})$ up to the first order.

Then we will derive that
\begin{equation} \label{temperature}
2 \pi T = \frac{\displaystyle \sqrt{2\kappa_{o}}\sqrt{r-r_{+}}}{\displaystyle\frac{1}{\sqrt{2\kappa_{o}}}\int^{r}_{r_{+}}\frac{1}{\sqrt{r'-r_{+}}}dr'} = \kappa_{o},
\end{equation}
and this will be true as long as $r$ approaches $r_{+}$. This completes the proof.

Second, we use the Parikh and Wilczek's tunneling method \cite{Parikh:1999mf}.
Although there are coordinate singularities around $r_{+}$, if we choose a good coordinate system (the Painleve-Gullstrand form),
we can regularize them \cite{Visser:2001kq}. Then the radial out-going null geodesics are given by
\begin{equation} \label{nullgeodesic}
\dot{r}\equiv\frac{dr}{dt}=1-\sqrt{\frac{2M(r)}{r}}.
\end{equation}
In this case, Parikh and Wilczek suggest \cite{Parikh:1999mf} that
\begin{equation} \label{Gamma}
\Gamma \sim e^{-2\mathrm{Im}S} \sim e^{-E/T},
\end{equation}
where $\Gamma$ is the emission rate, and $S$ is the action related to the tunneling.

Now the action is calculated as follows:
\begin{eqnarray} \label{integration1}
S = \int_{r_{in}}^{r_{out}}p_{r}dr = \int_{r_{in}}^{r_{out}}\int_{0}^{p_{r}}dp'_{r}dr\nonumber \\
 = \int_{M}^{M-\omega}\int_{r_{in}}^{r_{out}}\frac{dr}{\dot{r}}dH.
\end{eqnarray}
After changing $\dot{r}$ to (\ref{nullgeodesic}) and $H$ to $m-\omega$, we obtain an integral
\begin{equation} \label{integration2}
S = \int_{0}^{+\omega}\int_{r_{in}}^{r_{out}} \frac{dr}{1-\sqrt{\frac{2M(r)}{r}}}(-d\omega').
\end{equation}
Although $H=m-\omega$, for simplicity, we just use $m$ and ignore the back-reaction.
Now we change $r_{in}$(i.e., $r_{+}$) to $2M$, and we assume that $r_{out}$ is slightly smaller than $r_{in}$.
To evaluate the integration, we expand $M(r)/r$ around $r_{+}$, and we get
\begin{equation} \label{integration3}
\int_{0}^{+\omega}(-d\omega') \int_{r_{in}}^{r_{out}} \frac{dr}{1-\sqrt{1-2\kappa_{o}(r-r_{+})}}.
\end{equation}
If we change the variable $r$ to $r-r_{+}$, since $r_{out}-r_{+}$ is less than 0, the only possible imaginary term comes from $-\mathrm{Log}(r)/\kappa_{o}$.
Then the imaginary part is $\omega\pi/\kappa_{o}$. Finally we get the Hawking temperature $\kappa_{o} / 2\pi$. This completes the proof.

According to these proofs, now we obtain a reasonable formula for the Hawking temperature of regular black holes.

Now, let us think about the extreme case. In this case, since the condition is $F'(r_{\pm})=0$, the result is $M'(r_{\pm})=1/2$,
and this will give $T=0$.
If the black hole evolves slowly, then we can use this formula successively.
As time goes on, the black hole will approach the extreme case, but in this limit, the Hawking temperature approaches $0$.
Therefore, the final stage of the black hole can be assumed to slowly vary. Then we can write
\begin{equation} \label{dmdt}
\frac{dm}{dt} \sim -T^{\alpha},
\end{equation}
where $\alpha$ is a positive constant, and one can notice that the black hole will approach the extreme limit.
However, one can think that the thermodynamical description may be false if the situation is highly dynamic.
Anyway, although it is actually true and two horizons disappear completely, it will make the duplication experiment clearer.

To extend the thin shell approximation, one may suggest a regular $M$ function as $mA(r)+(r^{3}/2l^{2})B(r)$ where $A(r)$ ($B(r)$) begins with $1$ from the outside (inside) and quickly decreases to $0$ as the radius changes along $r_{0}$.
By using this method, we may modify and extend the metric around the extreme limit.

\section{\label{appb}False Vacuum Generation in a True Vacuum}

\begin{figure}
\begin{center}
\includegraphics[scale=0.6]{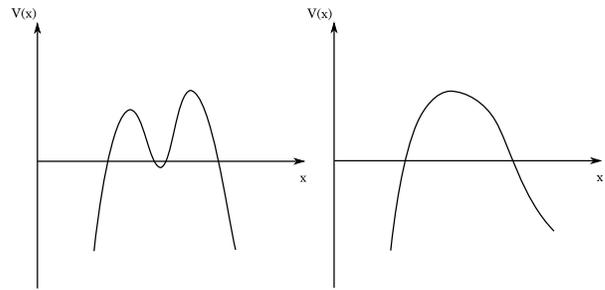}
\caption{\label{fig:potential}The potential-like function $V(x)$. The left one is for \cite{Frolov:1988vj}. The right one is for \cite{Blau:1986cw}.}
\end{center}
\end{figure}

We should consider two facts to make a local false vacuum.
One is the energy conservation problem, and the other is the stability of the mass shell.

If the background itself is a kind of de Sitter, since the energy cannot be defined globally and the scalar field can have thermal fluctuations,
a tunneling or roll-down process will be possible \cite{Lee:1987qc}.

Now, let us choose the metric form as
\begin{eqnarray} \label{metric_original}
ds^{2}=-\left(1-\frac{2m(r,l)}{r}-\frac{1}{3}\tilde{\Lambda}r^{2}\right) dt^{2}\nonumber \\
+\left(1-\frac{2m(r,l)}{r}-\frac{1}{3}\tilde{\Lambda}r^{2}\right)^{-1}dr^{2}+r^{2}d\Omega^{2},
\end{eqnarray}
where $\tilde{\Lambda}$ is a cosmological constant of the background.
As long as $1/M^{2}\gg\tilde{\Lambda}$, where $M$ is typical mass or length scale of our model (so $M \gtrsim l$),
the metric form will be similar to (\ref{metric}).
So the only difference of the causal structure is to change the past and future infinity to the cosmological horizon.

For the stability of mass shell, we use the method which is considered in \cite{Balbinot:1990}.
Define a parameter $x$ by
\begin{equation}
R=l\left(\frac{m}{l}\right)^{1/3} x,
\end{equation}
where $R$ is the location of the mass shell, and $m$ is the black hole mass.
The researchers argue that the parameter $x$ satisfies
\begin{equation}
\dot{x}^{2}+V(x)=a^{2}
\end{equation}
by a constant $a$ and some potential-like function $V(x)$ (Fig. \ref{fig:potential}).

We can see that the Frolov, Markov, and Mukhanov's model has a stable local minimum,
and this implies that the mass shell is stable during small fluctuations of parameter $m$ or $l$.
However, a large perturbation results in a significant change of $x$.
If the false vacuum decays (but not to $0$ for maintaining the regularity of the center),
$l$ will increase, and $x$ will become smaller and smaller.
This implies that the mass shell collapses and forms a singularity.

This technique was also considered by \cite{Blau:1986cw}\cite{tunneling}.
According to \cite{Balbinot:1990}, the potential-like function for \cite{Blau:1986cw} has no stable local minimum,
so the mass shell must be dynamic.
If its initial condition is nonsingular, it will behave in a left-rolling manner (i.e., $x$ will decrease)\cite{Farhi:1986ty}.
However, if we assume tunneling from left-rolling to right-rolling, or from a small vacuum to a large vacuum, a baby universe will be possible;
however, since this situation may imply the violation of unitarity in the context of the holographic principle \cite{Freivogel:2005qh}, this must be considered carefully.

\end{document}